\documentclass[12pt]{article}
\usepackage{epsf}
\begin{document}

\newcommand{\gtrsim}{ \mathop{}_{\textstyle \sim}^{\textstyle >} }
\newcommand{\lesssim}{ \mathop{}_{\textstyle \sim}^{\textstyle <} }

\newcommand{\rem}[1]{{\bf #1}}

\renewcommand{\theequation}{\thesection.\arabic{equation}}

\renewcommand{\thefootnote}{\fnsymbol{footnote}}
\setcounter{footnote}{0}
\begin{titlepage}

\def\thefootnote{\fnsymbol{footnote}}

\begin{center}

\hfill IASSNS-HEP-99-54\\
\hfill PUPT-1873\\
\hfill MIT-CTP-2873\\
\hfill hep-ph/9906527\\
\hfill June, 1999\\

\vskip .5in

{\Large \bf
  Wino Cold Dark Matter from Anomaly-Mediated SUSY Breaking
}

\vskip .45in

{\large
  Takeo Moroi$^{1,}$\footnote
  {Email: {\tt moroi@ias.edu}}
  and Lisa Randall$^{2,3,}$\footnote
  {Email: {\tt randall@feynman.princeton.edu}}
}

\vskip 0.2in

{\em ${}^{1}$
  School of Natural Sciences,
  Institute for Advanced Study, Princeton, NJ 08540, U.S.A.}

\vskip 0.1in

{\em ${}^{2}$
  Joseph Henry Laboratories, Princeton University,
  Princeton, NJ 08543, U.S.A.}

\vskip 0.1in

{\em ${}^{3}$
  Center for Theoretical Physics,
  Massachusetts Institute of Technology\\
  Cambridge, MA 02139, U.S.A.}

\end{center}

\vskip .4in

\begin{abstract}

The cosmological moduli problem is discussed in the framework of
sequestered sector/anomaly-mediated supersymmetry (SUSY) breaking.  In
this scheme, the gravitino mass (corresponding to the moduli masses)
is naturally 10 $-$ 100~TeV, and hence the lifetime of the moduli
fields can be shorter than $\sim {\rm 1~sec}$.  As a result, the
cosmological moduli fields should decay before big-bang
nucleosynthesis starts.  Furthermore, in the anomaly-mediated
scenario, the lightest superparticle (LSP) is the Wino-like
neutralino. Although the large annihilation cross section means the
thermal relic density of the Wino LSP is too small to be the dominant
component of cold dark matter (CDM), moduli decays can produce Winos
in sufficient abundance to constitute CDM. If Winos are indeed the
dark matter, it will be highly advantageous from the point of view of
detection. If the halo density is dominated by the Wino-like LSP, the
detection rate of Wino CDM in Ge detectors can be as large as $0.1 -
0.01$~event/kg/day, which is within the reach of the future CDM
detection with Ge detector. Furthermore, there is a significant
positron signal from pair annihilation of Winos in our galaxy which
should give a spectacular signal at AMS.

\end{abstract}
\end{titlepage}

\renewcommand{\thepage}{\arabic{page}}
\setcounter{page}{1}
\renewcommand{\thefootnote}{\#\arabic{footnote}}
\setcounter{footnote}{0}

\section{Introduction}
\setcounter{equation}{0}

In string theory, there are usually many flat directions that are
expected to acquire a mass from supersymmetry (SUSY) breaking.  The
mass is of order the supersymmetry breaking mass; the fields are very
light but have no collider implications because their interactions are
suppressed by the Planck scale.  However, moduli fields can be very
important (and dangerous) from a cosmological vantage
point~\cite{moduli_prob,PLB318-447}. The moduli fields are expected to
have a Planck scale amplitude in the early universe, and will
therefore dominate the energy density of the universe as soon as they
start to oscillate.  However, the modulus lifetime for a
Planck-coupled modulus field is so long that standard cosmological
scenarios are adversely affected~\cite{moduli_prob,PLB318-447}.  Most
important for generic moduli with mass of order the electroweak scale,
it can be shown that the moduli decay occurs after big-bang
nucleosynthesis (BBN), destroying the successful predictions. This
problem is referred to as the ``cosmological moduli problem.''

One potential resolution of this problem is that the gravitino mass
(i.e., the moduli mass) is larger than generically assumed; this
requires the mass to be 10 $-$ 100~TeV~\cite{PLB174-176,PLB342-105}.
With this mass, the modulus lifetime can be shorter than 1~sec, so
that decay occurs before BBN and the standard BBN is unaffected. In
``standard'' gravity-mediated scenarios, such a large modulus mass is
difficult to understand. Furthermore, raising the modulus mass
permitting a more rapid decay does not automatically lead to a
successful cosmology.  This is because the decay of a modulus field
can produce a sizable number of superparticles which cascade down to
the lightest superparticle (LSP).  Assuming $R$-parity conservation,
it has been claimed that the relic density of the LSP is likely to
overclose the
Universe~\cite{PLB342-105,NPB449-229,PLB370-52,PLB438-267}.

Recently, however, a novel framework for supersymmetry breaking has
been proposed in which SUSY breaking parameters are generated by the
super-Weyl anomaly effects~\cite{hth9810155,JHEP9812-027} and are
therefore loop-suppressed relative to the standard ``hidden sector''
predictions.  In particular, this scenario predicts the gaugino masses
as
 \begin{eqnarray}
  m_{{\rm G}i} = \frac{b_i g_i^2}{16\pi^2} \langle M\rangle,
 \end{eqnarray}
 where $g_i$ are the gauge coupling constants with $i=1,2,3$
identifying the gauge group, and $b_i$ are the $\beta$-function
coefficients of the gauge coupling constant.  Furthermore, $M$ is the
auxiliary field in the supergravity multiplet whose vacuum expectation
value (VEV) is expected to be of order the gravitino mass $m_{3/2}$.
The above relation tells us two important consequence of the
anomaly-mediated mass spectrum.  The first is that the Wino becomes
the LSP, instead of the Bino which is the conventional candidate for
the LSP.  Indeed, substituting the weak scale values of $g_i$, the
anomaly-mediated model predicts $m_{\rm G1}:m_{\rm G2}:m_{\rm
G3}\simeq 3:1:-10$.  Second, an important feature is that the
gravitino is extremely heavy in this framework.  Since the gaugino
masses are one-loop suppressed relative to the gravitino mass, the
gaugino masses at the electroweak scale require the gravitino mass to
be 10 $-$ 100~TeV.  In the scenario of Ref.~\cite{hth9810155},
``sequestered sector SUSY breaking,'' a consistent theory is presented
in which the scalar masses are sufficiently light, despite the large
gravitino mass.  Therefore, in this framework, namely sequestered
sector SUSY breaking with an anomaly-mediated mass spectrum for
gauginos, it is quite natural to expect the large gravitino mass that
solves the cosmological moduli problem.  Some alternative
possibilities for solving the moduli problem with a light modulus
mass, such as using an enhanced symmetry point~\cite{PRL75-398} or
late time inflation~\cite{NPB449-229,lateinf}, have also been
suggested.  However, the heavy modulus mass is probably the simplest
possibility.

Furthermore, the problem of too large a residual LSP mass density
(even in the presence of a large gravitino mass) assumed a Bino-like
neutralino whose pair annihilation cross section is $p$-wave
suppressed.  However, if the LSP is not Bino-like, the interaction of
the LSP changes and a larger pair annihilation cross section may be
realized.  This suppresses the mass density of the LSP.  With the
anomaly-mediated spectrum, the Wino-like neutralino is the LSP.  Since
the Wino has a larger pair annihilation cross section than the Bino,
the mass density of the LSP can be sufficiently suppressed.  It should
be also noted that the number of the LSP produced by the decay of one
modulus field is usually assumed to be $O(1)$.  However, the number of
produced LSP is model-dependent, so a much smaller number density of
the LSP could be produced, as we will discuss in Appendix.  Based on
these observations, we will demonstrate that not only can the
cosmological moduli problem be solved in the anomaly-mediated SUSY
breaking (AMSB) scenarios, but, furthermore, the LSP relic density
from modulus decay can be reduced to an acceptable level. In fact, if
the parameters are right, the Wino is a perfect dark matter candidate.

This has an important advantage from the point of view of detecting
SUSY dark matter.  Ordinarily, the thermal relic density and the
anti-matter fluxes are both determined by the strength of the dark
matter candidate's coupling.  A large detection efficiency requires a
large coupling, while a large relic density requires a relatively
small coupling in order to impede annihilation. This in general
implies relatively low efficiency for detecting supersymmetric dark
matter.  In our scenario however, because the Winos are produced from
moduli decay, there can be sufficiently many to comprise dark matter,
despite the large cross section. This is good both from the vantage
point of standard detection, and also for the new searches for
anti-matter, particularly by the Alpha Magnetic Spectrometer (AMS).

The outline of this paper is as follows.  We  calculate the mass
density of the Wino-like LSP produced by the decay of the moduli
fields in Section~\ref{sec:omega}.  We will see that the density
parameter $\Omega_\chi$ of the LSP can be in a reasonable range (0.1
$-$ 1) in some regions of parameter space, and hence the old problem
of the overproduction of the LSP due to the modulus decay may be
solved in AMSB scenario.  Furthermore, this fact gives us a motivation
to consider the relic Wino as the dominant component of  cold dark
matter (CDM).  In Section~\ref{sec:signal}, we discuss
possible signals from  Wino CDM. We will see that the larger Wino
cross sections permit a much more optimistic scenario for the
possibility of detecting SUSY dark matter than the more conventional
type.  In section~\ref{sec:discussion}, we conclude.  In
Appendix~\ref{app:property}, the properties of the moduli fields are
discussed.

\section{Mass Density of the Wino LSP}
\label{sec:omega}
\setcounter{equation}{0}

We first discuss the cosmological evolution of the modulus field and
the density of the LSP. In the very early Universe, the modulus field
has a large amplitude, expected to be as large as the Planck scale. It
begins to oscillate when the expansion rate $H$ of the Universe
becomes comparable to $m_\phi$.\footnote
 {Even if the initial amplitude of the modulus field is smaller than
$O(M_*)$, the energy density of the Universe is dominated by that of
the modulus field when $\phi$ decays if the initial amplitude is
larger than $\sim {\rm 10^{12}}$~GeV.  In this paper, we assume this
is the case.}
 After this period, the energy density of the Universe is dominated by
that of the modulus field. Then, when $H\sim\Gamma_\phi$, the modulus
field decays. The decay products are quickly thermalized and the
Universe is reheated. Furthermore, the decay of the modulus field
produces LSP's. Produced LSP's also lose their energy by scattering
off background particles and become non-relativistic.

The evolution of the number density $n_\chi$ of the LSP is obtained by
solving the following coupled Boltzmann equations:
 \begin{eqnarray}
  \frac{dn_\chi}{dt} + 3Hn_\chi
  &=& \bar{N}_{\rm LSP} \Gamma_\phi n_\phi
  - \langle v_{\rm rel}\sigma\rangle n_\chi^2,
 \label{dnchi/dt} \\
  \frac{dn_\phi}{dt} + 3Hn_\phi &=& -\Gamma_\phi n_\phi,
 \\
  \frac{d\rho_{\rm rad}}{dt} + 4H\rho_{\rm rad} &=&
  (m_\phi - \bar{N}_{\rm LSP} m_\chi) \Gamma_\phi n_\phi
  + 2 m_\chi \langle v_{\rm rel}\sigma\rangle n_\chi^2,
 \label{drho/dt}
 \end{eqnarray}
 where $m_\chi$ is the mass of the LSP, and $\bar{N}_{\rm LSP}$ is the
averaged number of the LSP produced in the decay of one modulus field.
Here, $n_\phi$ is the number density of the modulus field which is
related to the energy density of the modulus field $\rho_\phi$ as
$\rho_\phi=m_\phi n_\phi$. The quantity $\rho_{\rm rad}$ is the energy
density of the radiation which is related to the background
temperature $T$ as $\rho_{\rm rad} = \frac{\pi^2}{30}g_* T^4$, where
$g_*$ is the effective number of the massless degrees of freedom. In
our calculation, we use $g_*=10.75$, since we consider a situation
with a reheating temperature of $T_{\rm R}\sim O(1-10~\mbox{MeV})$.

One important quantity in solving these Boltzmann equations is the
thermally averaged annihilation cross section $\langle v_{\rm
rel}\sigma\rangle$. If the LSP is Bino-like, the annihilation is
through $p$-wave processes and $\langle v_{\rm rel}\sigma\rangle$ is
suppressed.  As a result, the decay of the modulus field overproduces
LSP's for a reasonable reheating temperature of $T_{\rm R}\sim
O(1-10~\mbox{MeV})$~\cite{PLB174-176,PLB342-105}.  However, in the
case of the Wino LSP, the pair annihilation proceeds through an
$s$-wave process. In particular, by exchanging a charged Wino, the
neutral Wino (i.e., LSP) can annihilate into a $W$-boson pair. In the
non-relativistic limit, the annihilation cross section is given by
 \begin{eqnarray}
  \langle v_{\rm rel}
  \sigma_{\tilde{W}^0\tilde{W}^0 \rightarrow W^+W^-} \rangle
  = \frac{g_2^4}{2\pi} \frac{1}{m_\chi^2}
  \frac{(1-x_W)^{3/2}}{(2-x_W)^{2}},
 \label{vs_WW}
 \end{eqnarray}
 where $x_W\equiv m_W^2/m_\chi^2$, and $g_2$ is the gauge coupling
constant of SU(2)$_{\rm L}$.  In the following calculation, we use
this formula for the annihilation cross section of the LSP.\footnote
 {We neglect the possible co-annihilation of charged and neutral
Winos.  If the Wino is in kinematic equilibrium, the number density of
the charged Wino is extremely suppressed.  This is because the mass
splitting between charged and neutral Winos is of order 100~MeV $-$
1~GeV~\cite{hph9904250,hph9904378}, which is much larger than the
temperature we are considering.  In this case, our approximation is
extremely well justified.}

Another important parameter is $\Gamma_\phi$, the decay width of the
modulus field.  Since the interaction of the modulus field is
proportional to inverse powers of $M_*$, where $M_*\simeq 2.4\times
10^{18}~\mbox{GeV}$ is the reduced Planck scale, $\Gamma_\phi$ is
extremely suppressed as seen in Appendix~\ref{app:property}.
 In order to discuss this in a model-independent way, we parameterize
the decay width as
 \begin{eqnarray}
  \Gamma_\phi = \frac{1}{2\pi} \frac{m_\phi^3}{\Lambda_{\rm eff}^2}.
 \end{eqnarray}
 Here, $\Lambda_{\rm eff}$ is the effective suppression scale for the
interaction of the modulus field, which can be determined as a
function of the couplings given in the previous section.  For an
unsuppressed two-body decay process, we expect $\Lambda_{\rm eff}\sim
M_*$.

It is instructive to discuss the qualitative behavior of the solution
to the Boltzmann equations (\ref{dnchi/dt}) $-$ (\ref{drho/dt}).
Since the modulus field decays when the expansion rate becomes
comparable to $\Gamma_\phi$, the reheating temperature is estimated as
 \begin{eqnarray}
  T_{\rm R} \sim \left(\frac{\pi^2}{90}g_*\right)^{-1/4}
  \sqrt{\Gamma_\phi M_*}
  \sim 7.7~\mbox{MeV} \times
  \left(\frac{m_\phi}{100~\mbox{TeV}}\right)^{3/2}
  \left(\frac{\Lambda_{\rm eff}}{M_*}\right)^{-1}
  \left(\frac{g_*}{10.75}\right)^{1/4},
 \label{T_R}
 \end{eqnarray}
 where we used the instantaneous decay
approximation~\cite{Kolb-Turner}.  This reheating temperature has to
be larger than $\sim 1~{\rm MeV}$ in order not to affect the success
of standard big-bang nucleosynthesis~\cite{PRL82-4168}.  This is
guaranteed for a modulus mass of $\sim 100~{\rm TeV}$ with a naive
two-body decay rate
$\Gamma_\phi\sim\frac{1}{4\pi}\frac{m_\phi^3}{M_*^2}$.  So, from the
vantage point of the standard cosmological moduli problem, the
sequestered sector scenario is very advantageous.  In order to achieve
this estimated two-body decay rate, the moduli should decay into gauge
boson pairs, Higgs pairs, or gravitino pairs through the interactions
given in Eqs.~(\ref{L_G}), (\ref{L_H}), or (\ref{L_psi}),
respectively, with unsuppressed coupling constant (i.e., $\lambda\sim
1$).

With the decays of the moduli fields, the LSP is produced.  However,
the evolution of the cosmological density of the LSP is different from
the usual case.  In the standard scenario, the produced LSP's reach
thermal equilibrium.  Therefore, when the temperature $T$ is higher
than $m_\chi$, the number density of the LSP is comparable to those of
massless particles, while once $T$ becomes lower than $m_\chi$,
$n_\chi$ is Boltzmann suppressed.  Then, at some temperature $T_{\rm
dec}$, the pair annihilation rate of the LSP becomes smaller than the
expansion rate of the Universe, and the LSP decouples from the thermal
bath; after this stage, the number of the LSP in a comoving volume is
fixed.  In this case, the thermal relic density of the Wino-like LSP
is estimated as~\cite{JHEP9812-027}
 \begin{eqnarray}
  \Omega_\chi^{\rm (thermal)} h^2 \simeq
  5\times 10^{-4} \times \left(\frac{m_\chi}{\rm 100~GeV}\right)^2,
 \label{Omega_th}
 \end{eqnarray}
 where $h$ is the present Hubble constant in units of 100~km/sec/Mpc.
Obviously, in the standard scenario, the Wino LSP cannot be the
dominant component of the CDM since the mass density given above is
too small.

If $T_{\rm R}$ is higher than $T_{\rm dec}$, the relic density of the
Wino-like LSP is given by Eq.~(\ref{Omega_th}).  However, the typical
decoupling temperature is given by $T_{\rm dec}\sim
\frac{1}{30}m_\chi$~\cite{Kolb-Turner}, and hence Eq.~(\ref{T_R})
tells us that $T_{\rm R}$ is much lower than $T_{\rm dec}$.  In this
case, the LSP from the modulus decay is never in chemical equilibrium,
and its number density just decreases because of pair annihilation.
However, the pair annihilation rate eventually becomes smaller than
the expansion rate of the Universe, since the annihilation rate is
proportional to the number density of the LSP.  Once this happens, the
pair annihilation is no longer effective, and the LSP freezes out.
This happens when the annihilation term in Eq.~(\ref{dnchi/dt}) (i.e.,
the second term in RHS) becomes less significant than the dilution
term (i.e., the second term in LHS).  The number density is estimated
as
 \begin{eqnarray}
  n^{({\rm ann})}_\chi(T_{\rm R})
  \sim
  \left.
  \frac{3H}{2\langle v_{\rm rel}\sigma\rangle}
  \right|_{T=T_{\rm R}}
  \sim
  \frac{3\Gamma_\phi}{2\langle v_{\rm rel}\sigma\rangle}.
 \end{eqnarray}
 The pair annihilation proceeds as long as $n_\chi$ is larger than
$n^{({\rm ann})}_\chi$ given above, and hence $n^{({\rm ann})}_\chi$
is the upper bound on the number density of the LSP for a given
$T_{\rm R}$.  It is notable that this upper bound is insensitive to
the mechanism for LSP production.  With this relation, we obtain the
mass density to entropy ratio as
 \begin{eqnarray}
  \frac{\rho^{({\rm ann})}_\chi}{s}
  &\sim&
  \frac{m_\chi n^{({\rm ann})}_\chi(T_{\rm R})}
  {(2\pi^2/45)g_*T_{\rm R}^3}
 \nonumber \\
  &\sim&
  1.3\times 10^{-9}~\mbox{GeV}\times
  \frac{(2-x_W)^{2}}{(1-x_W)^{3/2}}
 \nonumber \\ && \times
  \left(\frac{m_\chi}{100~\mbox{GeV}}\right)^3
  \left(\frac{m_\phi}{100~\mbox{TeV}}\right)^{-3/2}
  \left(\frac{\Lambda_{\rm eff}}{M_*}\right)
  \left(\frac{g_*}{10.75}\right)^{-1/4}.
 \label{rho/s_ann}
 \end{eqnarray}
 Since we expect no entropy production after the decay of $\phi$, the
above ratio should be preserved until today.  One can easily see that
the ratio $\rho^{({\rm ann})}_\chi/s$ is proportional to $T_{\rm
R}^{-1}$.  For $T_{\rm R}\sim T_{\rm dec}$, Eq.~(\ref{rho/s_ann})
approximately reproduces the standard result given in
Eq.~(\ref{Omega_th}).  However, since the reheating temperature is
much lower than $T_{\rm dec}$, we expect a significantly larger mass
density of the LSP as a result of moduli decay.  One should note that
the above ratio is proportional to $\Lambda_{\rm eff}$, and that the
mass density becomes smaller as the modulus field interacts more
strongly.

The result given in Eq.~(\ref{rho/s_ann}) is valid only if there is a
sufficiently large number of LSP's produced by the decay of the
modulus field.  If there is insufficient production, the pair
annihilation is not effective, and all the produced LSP's survive.  In
this case, the number density of LSP's is estimated as
 \begin{eqnarray}
  n^{(0)}_\chi(T_{\rm R})
  \sim \bar{N}_{\rm LSP} n_\phi(T_{\rm R})
  \sim \frac{3\bar{N}_{\rm LSP}\Gamma_\phi^2M_*^2}{m_\phi},
 \end{eqnarray}
 and hence
 \begin{eqnarray}
  \frac{\rho^{(0)}_\chi}{s}
  &\sim&
  \frac{m_\chi n^{(0)}_\chi(T_{\rm R})}{(2\pi^2/45)g_*T_{\rm R}^3}
 \nonumber \\
  &\sim&
  5.8\times 10^{-6}~\mbox{GeV}
 \nonumber \\ &&
  \times
  \bar{N}_{\rm LSP}
  \left(\frac{m_\chi}{100~\mbox{GeV}}\right)
  \left(\frac{m_\phi}{100~\mbox{TeV}}\right)^{1/2}
  \left(\frac{\Lambda_{\rm eff}}{M_*}\right)^{-1}
  \left(\frac{g_*}{10.75}\right)^{1/4}.
 \label{rho/s_0}
 \end{eqnarray}
 Notice that the mass density is proportional to $\bar{N}_{\rm LSP}$,
the average number of LSP's produced by one modulus decay.

Using Eqs.~(\ref{rho/s_ann}) and (\ref{rho/s_0}), the actual mass
density is estimated as
 \begin{eqnarray}
  \frac{\rho_\chi}{s}\sim
  \mbox{min}\left(
  \frac{\rho^{(0)}_\chi}{s},
  \frac{\rho^{({\rm ann})}_\chi}{s}
  \right).
 \end{eqnarray}
 By comparing the above quantity with the current critical density
 \begin{eqnarray}
  \frac{\rho_{\rm c}}{s} \simeq
  3.6 \times 10^{-9} {\rm ~GeV} \times h^2,
 \label{rhoc2s}
 \end{eqnarray}
 we obtain the density parameter $\Omega_\chi\equiv
\rho_\chi/\rho_{\rm c}$.

A more accurate estimation of the density parameter is given by
solving the Boltzmann equations.  We solved the Boltzmann equations
(\ref{dnchi/dt}) $-$ (\ref{drho/dt}) numerically and calculated the
density parameter $\Omega_\chi$ as a function of $\bar{N}_{\rm LSP}$,
$m_\phi$, $m_\chi$, and $\Lambda_{\rm eff}$. In our calculation, we
followed the evolution from the modulus-dominated era to the
radiation-dominated Universe where the temperature is much lower than
$T_{\rm R}$ and the LSP has already frozen out.

 \begin{figure}
 \centerline{\epsfxsize=0.5\textwidth\epsfbox{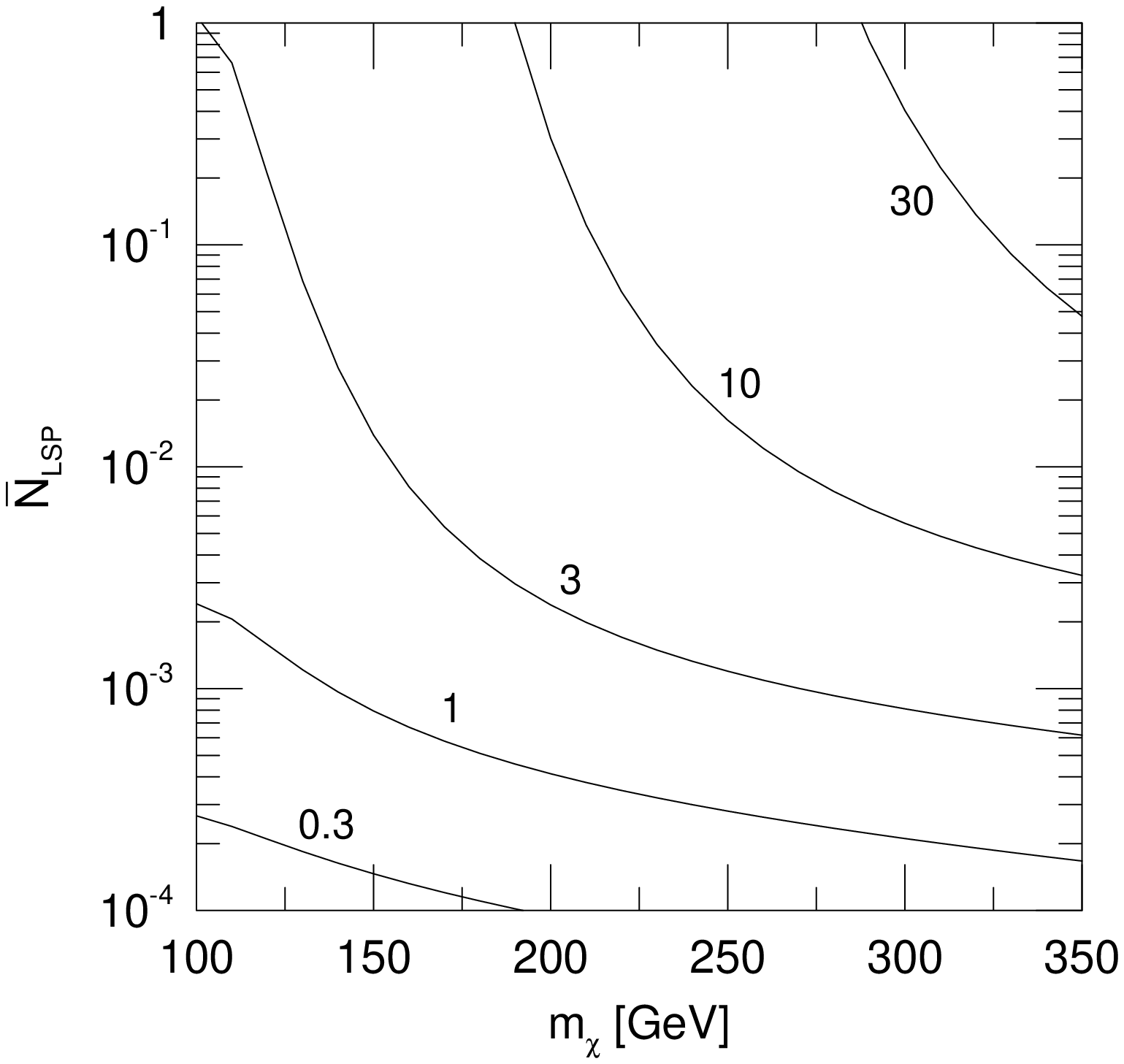}}
 \caption{Contours of the constant $\Omega_\chi h^2$ on $m_\chi$
vs.~$\bar{N}_{\rm LSP}$ plane.  Numbers in the figure are $\Omega_\chi
h^2$.  Here, we take $\Lambda_{\rm eff}=M_*$ and
$m_\phi=100~\mbox{TeV}$.}
 \label{fig:omg100}
 \vskip 1.5cm
 \centerline{\epsfxsize=0.5\textwidth\epsfbox{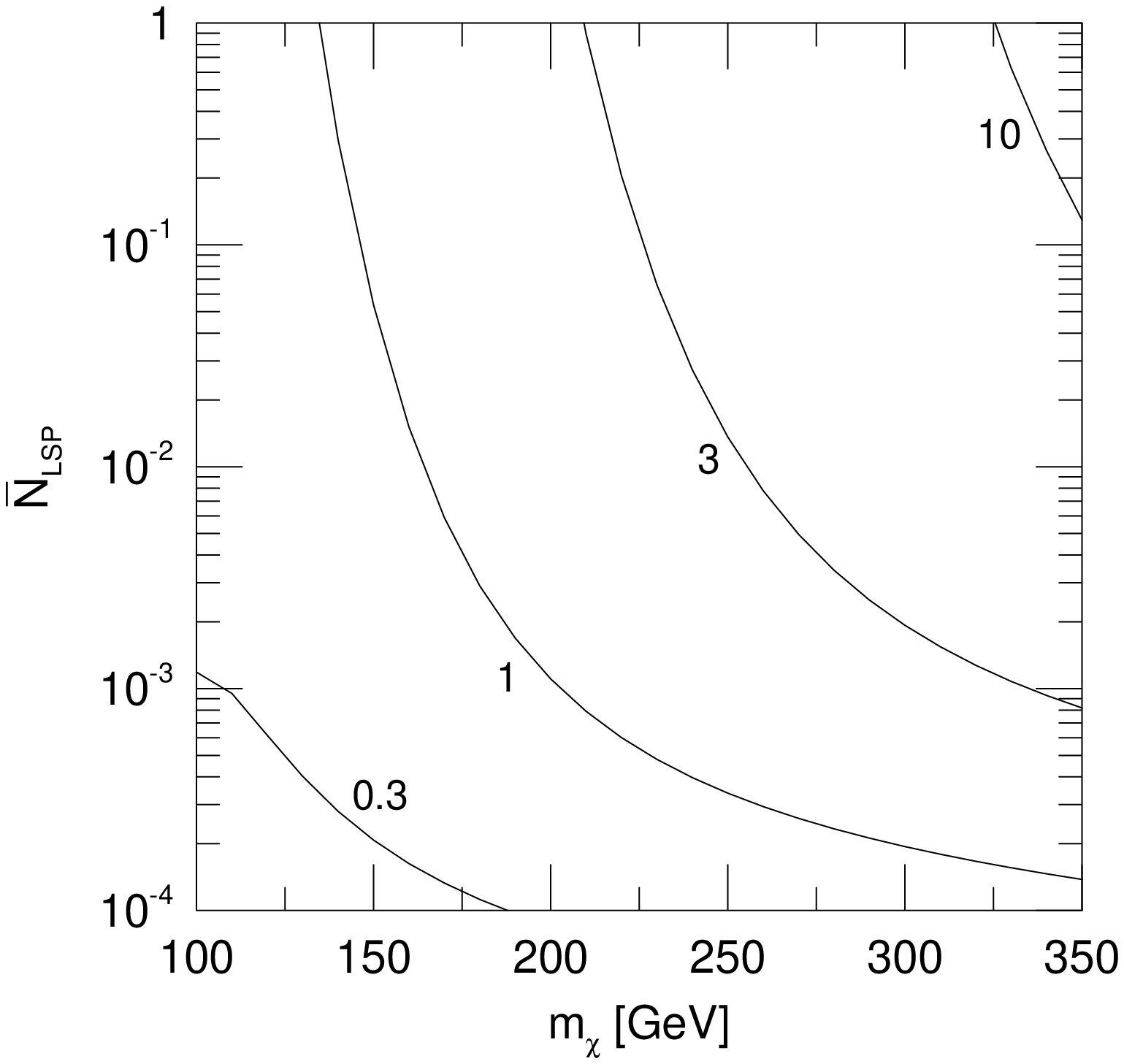}}
 \caption{Same as Fig.~\protect\ref{fig:omg100}, except
$m_\phi=300~\mbox{TeV}$.}
 \label{fig:omg300}
 \end{figure}

In Figs.~\ref{fig:omg100} and \ref{fig:omg300}, we plot the constant
$\Omega_\chi h^2$ contour as a function of $m_\chi$ and ~$\bar{N}_{\rm
LSP}$ for $m_\phi=100~\mbox{TeV}$ and 300~TeV respectively.  As we can
see, for $\bar{N}_{\rm LSP}\sim 1$, $\Omega_\chi h^2$ is almost
independent of $\bar{N}_{\rm LSP}$ since the pair annihilation is
important in this region.  (See Eq.~(\ref{rho/s_ann}).)  However, for
smaller $\bar{N}_{\rm LSP}$, the pair annihilation process becomes
ineffective and $\Omega_\chi h^2$ becomes sensitive to $\bar{N}_{\rm
LSP}$. (See Eq.~(\ref{rho/s_0}).)

With the natural value of $\Lambda_{\rm eff}\sim M_*$, $\Omega_\chi
h^2$ can be 0.1 $-$ 10, which is smaller than the result of the
conventional calculation with the Bino LSP.  In the AMSB, the Wino is
the LSP, and hence the pair annihilation among the LSP's is more
enhanced than the Bino LSP case.  Furthermore, the LSP density can be
significantly suppressed if $\bar{N}_{\rm LSP}\ll 1$.  (Notice that
the second effect has not been considered before, and it can be
important even in the Bino LSP case.)  Because of these two reasons,
the AMSB scenario realizes a smaller mass density of LSP's from moduli
decay.  In particular, $\Omega_\chi h^2$ can be 0.1 $-$ 1 almost
irrespective of the mass of the LSP for $\bar{N}_{\rm LSP}\sim
10^{-3}-10^{-4}$.  In this case, we can avoid the problem of
overclosing the Universe.  Furthermore, even with $\bar{N}_{\rm
LSP}\sim 1$, $\Omega_\chi h^2\sim O(1)$ if the Wino mass is less than
about 200~GeV.  So not only do we avoid the overproduction of Binos
which is part of the cosmological moduli problem, but we actually see
the the Wino is an excellent dark matter candidate.

The success of this scenario depends on the precise structure of the
operators through which the moduli decay.  The possible operators are
listed in Appendix~\ref{app:property}.
 If the moduli are lighter than twice the gravitino mass, the operator
(\ref{L_H}) should be the most important.  The operator (\ref{L_G})
might also be important; however its coefficient is very likely
loop-suppressed.  In either case, we would expect $\bar N_{\rm LSP}
\sim 10^{-3}-10^{-4}$,\footnote
 {If the decay is to Higgses, one would obtain the LSP's from the
suppressed three-body decay for example.}
 which is optimal for obtaining critical density for a large range of
Wino mass.  However, if $F_\phi$ is large, both operators (\ref{L_G})
and (\ref{L_H}) would have small coefficients in order to avoid too
large a gaugino mass and $\mu$-parameter in the sequestered sector
scenario. In this case, the moduli must decay into gravitino pairs
through the operator (\ref{L_psi}). This would give a sufficiently
high reheating temperature to avoid the first of the cosmological
moduli problems, but requires a small Wino mass to obtain critical
density (and not more) for the relic Winos.  In this case, one
requires slightly heavier moduli fields.

Finally, let us briefly discuss gravitino cosmology in the AMSB
scenario.  In the gravity-mediated SUSY breaking scenario, it is well
known that a gravitino much lighter than ${\rm 10~TeV}$ is
problematic~\cite{PRL48-1303}. In particular, the gravitinos are
produced in the early Universe through scattering processes, and their
decay can spoil the great success of the standard BBN scenario if
$m_{3/2}\ll {\rm 10~TeV}$~\cite{grav_bbn}.  Furthermore, the LSP may
be overproduced via the decay of the gravitino.  However, in the
sequestered sector case, these problems may be avoided since the
gravitino behaves almost like the modulus field.  In the sequestered
sector scenario, the gravitino mass can be so large that the decay
happens before BBN starts.  Furthermore, the LSP is produced by the
decay of the gravitino with $\bar{N}_{\rm LSP}\simeq 1$, but the
density of the LSP may be reduced by pair annihilation.  The only
difference is that the primordial number density of the gravitino is
determined by the reheating temperature just after the primordial
inflation, which we call $T_{\rm R}^{\rm (inf)}$.\footnote
 {$T_{\rm R}^{\rm (inf)}$ should be distinguished from $T_{\rm R}$,
the reheating temperature just after the decay of the modulus field.}
 If we can neglect the pair annihilation of the produced LSP, the
cosmological abundance of the LSP is proportional to the primordial
abundance of the gravitino.  In this case, if there is no effect from
the modulus field, the mass density of the LSP is given by
 \begin{eqnarray}
  \frac{\rho^{({\rm grav})}_\chi}{s} \simeq
  1.4 \times 10^{-9} \times
  \left(\frac{m_\chi}{100~\mbox{GeV}}\right)
  \left(\frac{T_{\rm R}^{\rm (inf)}}{10^{11}~\mbox{GeV}}\right),
 \label{rho/s_grav}
 \end{eqnarray}
 where we used the gravitino abundance given in Ref.~\cite{PTP93-879}.
However, once the number density becomes large enough, the pair
annihilation becomes effective and the above formula is not valid any
more.  In this case, it is relevant to use $\rho^{({\rm ann})}_\chi
/s$ given in Eq.~(\ref{rho/s_ann}), since this is the maximally
allowed mass density.  (For the gravitino case, in
Eq.~(\ref{rho/s_ann}), $\Lambda_{\rm eff}\sim M_*$ and $m_\phi
=m_{3/2}$.)  Therefore, the mass density of the LSP in this case is
estimated as
 \begin{eqnarray}
  \frac{\rho_\chi}{s}\sim
  \mbox{min}\left(
  \frac{\rho^{({\rm grav})}_\chi}{s},
  \frac{\rho^{({\rm ann})}_\chi}{s}
  \right).
 \end{eqnarray}
 When $T_{\rm R}^{\rm (inf)}\gtrsim 10^{11} - 10^{12}~{\rm GeV}$, the
pair annihilation is effective and the mass density of the LSP becomes
insensitive to $T_{\rm R}^{\rm (inf)}$.  In particular, if $m_\chi\sim
{\rm 100~GeV}$, $\Omega_\chi$ is always $O(1)$ for high enough $T_{\rm
R}^{\rm (inf)}$, and the Wino-like LSP can be a candidate for CDM.  If
$m_\chi\gg {\rm 100~GeV}$, $T_{\rm R}^{\rm (inf)}$ should be tuned to
be $\sim 10^{11}~{\rm GeV}$ to have $\Omega_\chi\sim 1$.  Of course,
if both the modulus field and the gravitino exist in the early
Universe, primordial gravitinos can be diluted by the decay of the
modulus field.  In this case, the mass density of the LSP produced by
the gravitino decay is negligible.  The gravitino cosmology is also
discussed in Ref.~\cite{hph9904378}, but the authors neglected the
effect of the pair annihilation of the LSP's.  As a result, they
claimed that $T_{\rm R}^{\rm (inf)}\sim {\rm 10^{11}~GeV}$ is
necessary to realize the Wino CDM.  Based on Eq.~(\ref{rho/s_grav}),
they also derived an upper bound on the reheating temperature of
$T_{\rm R}^{\rm (inf)}\lesssim {\rm 10^{11}~GeV}$ in order not to
overclose the Universe.  However, these arguments are modified as
above once we include the effect of pair annihilation.

Since the decay of the modulus field produces a large entropy, one may
worry about the dilution of the baryon asymmetry of the Universe in
our scenario.  It is true that the dilution factor due to the modulus
decay can be as large as $\sim 10^{13}$, and hence large primordial
baryon asymmetry is required.  However, this is not necessarily a
problem.  For example, the Affleck-Dine baryogenesis~\cite{NPB249-361}
can provide enough baryon number asymmetry even with such a large
dilution factor~\cite{PLB342-105}.

In summary, the mass density of the LSP produced by the decay of the
moduli fields can be sufficiently suppressed in the Wino-like LSP
case, and the AMSB scenario provides an interesting solution to the
cosmological moduli problem.  Furthermore, in this case, the Wino-like
LSP is a natural candidate for CDM. In the next section, we will see
this is a very advantageous situation from the point of view of dark
matter detection.

\section{Detecting Wino CDM}
\label{sec:signal}
\setcounter{equation}{0}

As we have seen, the Wino LSP is a promising candidate for cold dark
matter. It comes out quite naturally from the parameters of the
sequestered sector scenario.  In this section, we consider the
possibility of dark matter detection with Wino dark matter.

We first discuss the CDM search with Ge detectors.  If the LSP is the
dominant component of the mass density of the halo, we may observe the
energy deposit due to the LSP-nucleus scattering in a Ge detector.  In
the sequestered sector scenario, squark masses are calculable and are
generally quite heavy.  Therefore, the scattering processes mediated
by the squark exchange are suppressed.

However, the LSP $\chi$ also couples to the Higgs bosons as
 \begin{eqnarray}
  {\cal L}_{h\chi\chi} =
  y_{h\chi\chi} h\bar{\chi}\chi + y_{H\chi\chi} H\bar{\chi}\chi,
 \end{eqnarray}
 where $h$ and $H$ are the light and heavy CP even Higgses,
respectively.  Therefore, the LSP interacts with nuclei by exchanging
the Higgs bosons.  As we will see, this effect is important, and the
detection rate in the Ge detector can be as large as $\sim 0.1 -
0.01$~event/kg/day which should be within the reach of the future
detection of the CDM~\cite{PRep267-195}.

The above Yukawa coupling constants, $y_{h\chi\chi}$ and
$y_{H\chi\chi}$, arise from the mixings between the Wino and
Higgsinos.  These mixings are from the off-diagonal elements in the
neutralino mass matrix, which is given by
 \begin{eqnarray}
  {\cal M} =
  \left( \begin{array}{cccc}
  m_{\rm G1}  & 0
  & -m_W \cos\beta \tan\theta_{\rm W}
  & m_W \sin\beta \tan\theta_{\rm W} \\
  0 & m_{\rm G2}
  & m_W \cos\beta & -m_W \sin\beta \\
  -m_W \cos\beta \tan\theta_{\rm W}  & m_W \cos\beta
  & 0 & -\mu \\
  m_W \sin\beta \tan\theta_{\rm W} & -m_W \sin\beta
  & -\mu & 0
  \end{array} \right),
 \end{eqnarray}
 in the basis $(i\tilde{B},i\tilde{W}^0, \tilde{H}^0_1,
\tilde{H}^0_2)$.  Here, $m_{\rm G1}$ and $m_{\rm G2}$ are the gaugino
masses for the gauginos associated with the U(1)$_{\rm Y}$ and
SU(2)$_{\rm L}$ gauge groups respectively, $\mu$ is the supersymmetric
Higgs mass, $\tan\beta$ is the ratio of Higgs VEVs, and $\theta_{\rm
W}$ is the Weinberg angle.  In the AMSB scenario, $m_{\rm G1}$ and
$m_{\rm G2}$ are related by $m_{\rm G1}\simeq 3m_{\rm G2}$, which we
now assume.  The above mass matrix can be diagonalized by using a
unitary matrix, which we call $U$.\footnote
 {In our notation, the mass eigenstates of the neutralino are given by
$\chi_i = U_{1i}(i\tilde{B}) + U_{2i}(i\tilde{W}^0) +
U_{3i}\tilde{H}^0_1 + U_{4i}\tilde{H}^0_2$.}
 With this unitary matrix, the mass of the LSP is given by $m_\chi
=|(U^\dagger{\cal M}U)_{11}|$, for example.

The Yukawa coupling constants are given by
 \begin{eqnarray}
  y_{h\chi\chi} &=& -\frac{1}{2} g_2 U_{21}
  (U_{31}\cos\beta - U_{41}\sin\beta),
 \\
  y_{H\chi\chi} &=& -\frac{1}{2} g_2 U_{21}
  (U_{31}\sin\beta + U_{41}\cos\beta).
 \end{eqnarray}
 Here, we neglect the difference between the neutral CP even Higgs
mixing angle and the angle $\beta$.  In the decoupling limit (i.e.,
$m_h\ll m_H$), this difference is sufficiently small.  When $|\mu|$
and $|m_{\rm G2}|$ are much larger than $m_W$, the LSP is mostly the
Wino and $U_{21}\simeq 1$.  Other smaller elements are given by
 \begin{eqnarray}
  U_{11}\simeq 0,
  ~~~
  U_{31} \simeq
  -\frac{m_W (m_{\rm G2}\cos\beta - \mu\sin\beta)}
        {\mu^2 - m_{\rm G2}^2},
  ~~~
  U_{41} \simeq
  \frac{m_W (m_{\rm G2}\sin\beta + \mu\cos\beta)}
       {\mu^2 - m_{\rm G2}^2},
 \end{eqnarray}
 and in the same limit, $y_{h\chi\chi}$ and $y_{H\chi\chi}$ are given
by
 \begin{eqnarray}
  y_{h\chi\chi} &\simeq&
  \frac{g_2m_W(m_{\rm G2}+\mu\sin 2\beta)}{2(\mu^2-m_{G2}^2)},
 \label{y_h1xx}
 \\
  y_{H\chi\chi} &\simeq&
  -\frac{g_2m_W \mu\cos 2\beta}{2(\mu^2-m_{G2}^2)}.
 \end{eqnarray}
 As one can see, the coupling of the light Higgs is sensitive to the
relative sign of $m_{\rm G2}$ and $\mu$; $y_{h\chi\chi}$ is enhanced
if $\mu/m_{\rm G2}$ is positive.  Furthermore, as the ratio
$|\mu/m_{\rm G2}|$ becomes larger, the Higgsino component of the LSP
becomes smaller.  As a result, the $h\chi\chi$ and $H\chi\chi$
interactions are suppressed.  In the Bino LSP case, $y_{h\chi\chi}$
and $y_{H\chi\chi}$ are given by similar expressions with $g_2m_W$
being replaced by $g_1m_W\tan\theta_{\rm W}$.  Therefore, the Wino LSP
has stronger couplings to the Higgs boson than the Bino, which
enhances the detection cross section.

The interaction between the Higgs bosons and a nucleus are discussed
in Ref.~\cite{PLB78-443}.  Since we are interested in a scattering
process with small recoil energy, the nucleus can be approximately
regarded as an elementary particle with mass $m_N$.  Furthermore,
since $m_N$ is dynamically generated through QCD effects, $m_N$ is
proportional to the QCD scale $\Lambda_{\rm QCD}$.  On the other hand,
$\Lambda_{\rm QCD}$ is sensitive to the fluctuation of the Higgs
fields through the heavy quark mass dependence of $\Lambda_{\rm QCD}$.
By using these facts, we can relate the Higgs VEV dependence of the
QCD scale to the Higgs-nucleus coupling constants.  The interaction
terms of the Higgs bosons with the nucleus $N$ are derived
as~\cite{PLB78-443}
 \begin{eqnarray}
  {\cal L}_{\rm int} =
  \frac{\partial {\cal L}_{\rm mass}}{\partial m_N}
  (y_{hNN} h + y_{HNN} H),
 \end{eqnarray}
 where ${\cal L}_{\rm mass}$ is the mass term for the nucleus $N$ in
the effective theory.  We consider the case with Ge detector, and we
will use the formula for $N={\rm ^{76}Ge}$.  The ``effective''
coupling constants for the Higgs interactions are given by\footnote
 {We neglect the effects of the nuclear form factors.}
 \begin{eqnarray}
  y_{hNN} &=& -\frac{2}{9} \frac{m_N}{v},
 \\
  y_{HNN} &=& -\frac{2}{27} \frac{m_N}{v} (\tan\beta - 2 \cot\beta),
 \label{y_h2nn}
 \end{eqnarray}
 with $v\simeq {\rm 246~GeV}$.  As one can see in Eq.~(\ref{y_h2nn}),
the interaction of the heavy Higgs becomes stronger in the large
$\tan\beta$ case, and the heavy Higgs exchange diagram may enhance the
detection rate.

With the above coupling constants, the total cross section for the
scattering process $\chi N\rightarrow \chi N$ is given by
 \begin{eqnarray}
  \sigma_{\rm scatt} = \frac{4}{\pi}
  \left( \frac{y_{hNN}y_{h\chi\chi}}{m_h^2}
  + \frac{y_{HNN}y_{H\chi\chi}}{m_H^2} \right)^2
  \frac{m_N^2 m_\chi^2}{(m_N+m_\chi)^2},
 \end{eqnarray}
 where $m_h$ and $m_H$ are the masses of $h$ and $H$, respectively.
Assuming that the CDMs are virialized in the halo, we obtain the
detection rate
 \begin{eqnarray}
  R = \frac{2}{\sqrt{3\pi}}
  \frac{\rho_\chi^{\rm (halo)}\bar{v}_\chi\sigma_{\rm scatt}}{m_Nm_\chi}
  \exp\left(
  -\frac{3(m_N+m_\chi)^2E_{\rm thr}}{4m_Nm_\chi^2\bar{v}_\chi^2}
  \right),
 \label{Rate}
 \end{eqnarray}
 where $\rho_\chi^{\rm (halo)}$ is the mass density of the LSP in the
halo, $E_{\rm thr}$ is the threshold energy of the detector, and
$\bar{v}_\chi$ is the averaged velocity of the LSP in the halo.

 \begin{figure}
 \centerline{\epsfxsize=0.5\textwidth\epsfbox{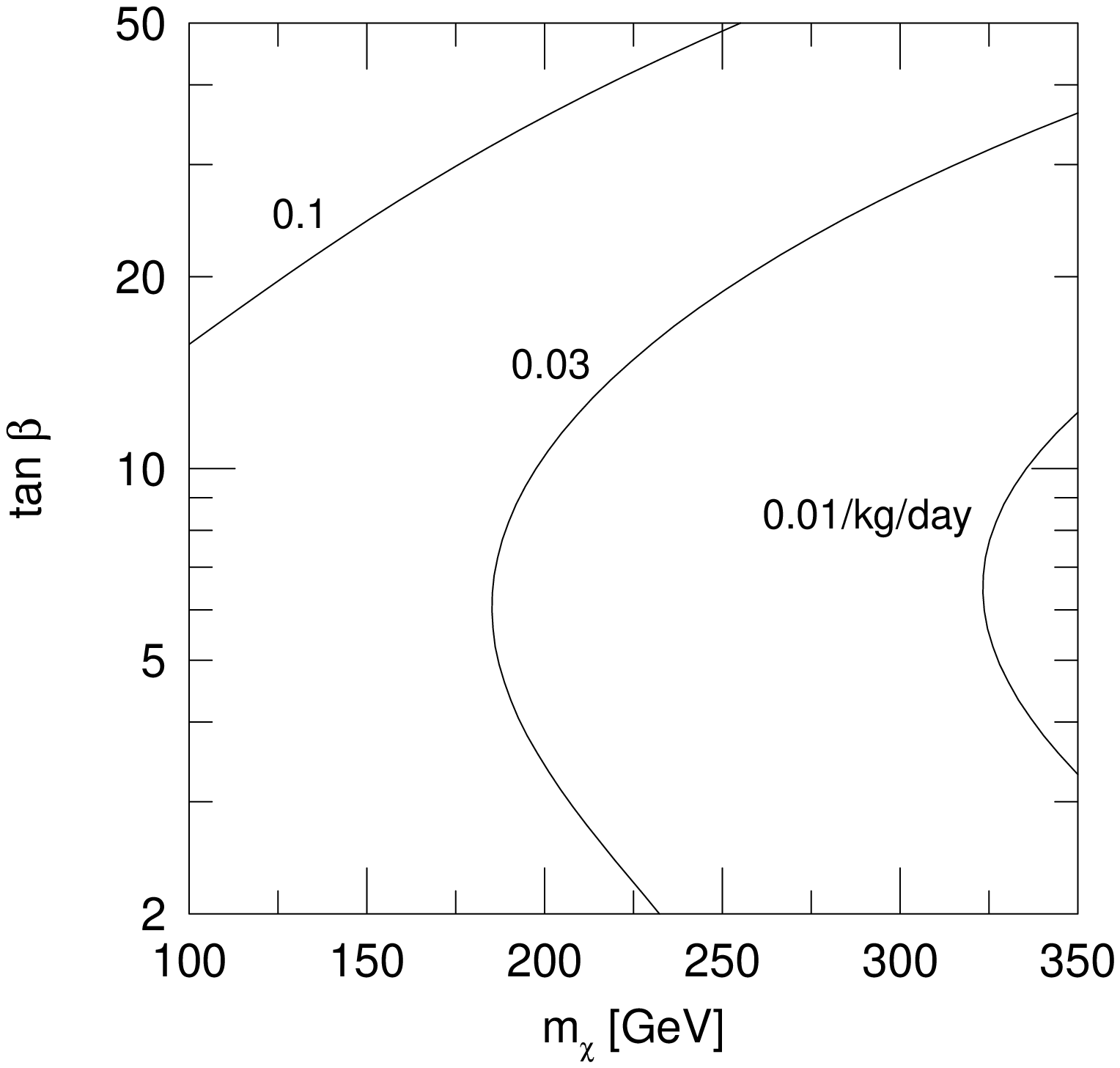}}
 \caption{Contours of the constant detection rate $R$ in $^{76}$Ge
detector on $m_\chi$ vs.~$\tan\beta$ plane.  We take the parameters of
$\mu =2m_{\rm G2}$, $m_h={\rm 100~GeV}$, $m_H={\rm 300~GeV}$, $E_{\rm
thr}={\rm 2~keV}$, $\bar{v}_\chi={\rm 320~km/sec}$ and $\rho_\chi^{\rm
(halo)}={\rm 0.3~GeV/cm^3}$.}
 \label{fig:R2}
 \vskip 1.5cm
 \centerline{\epsfxsize=0.5\textwidth\epsfbox{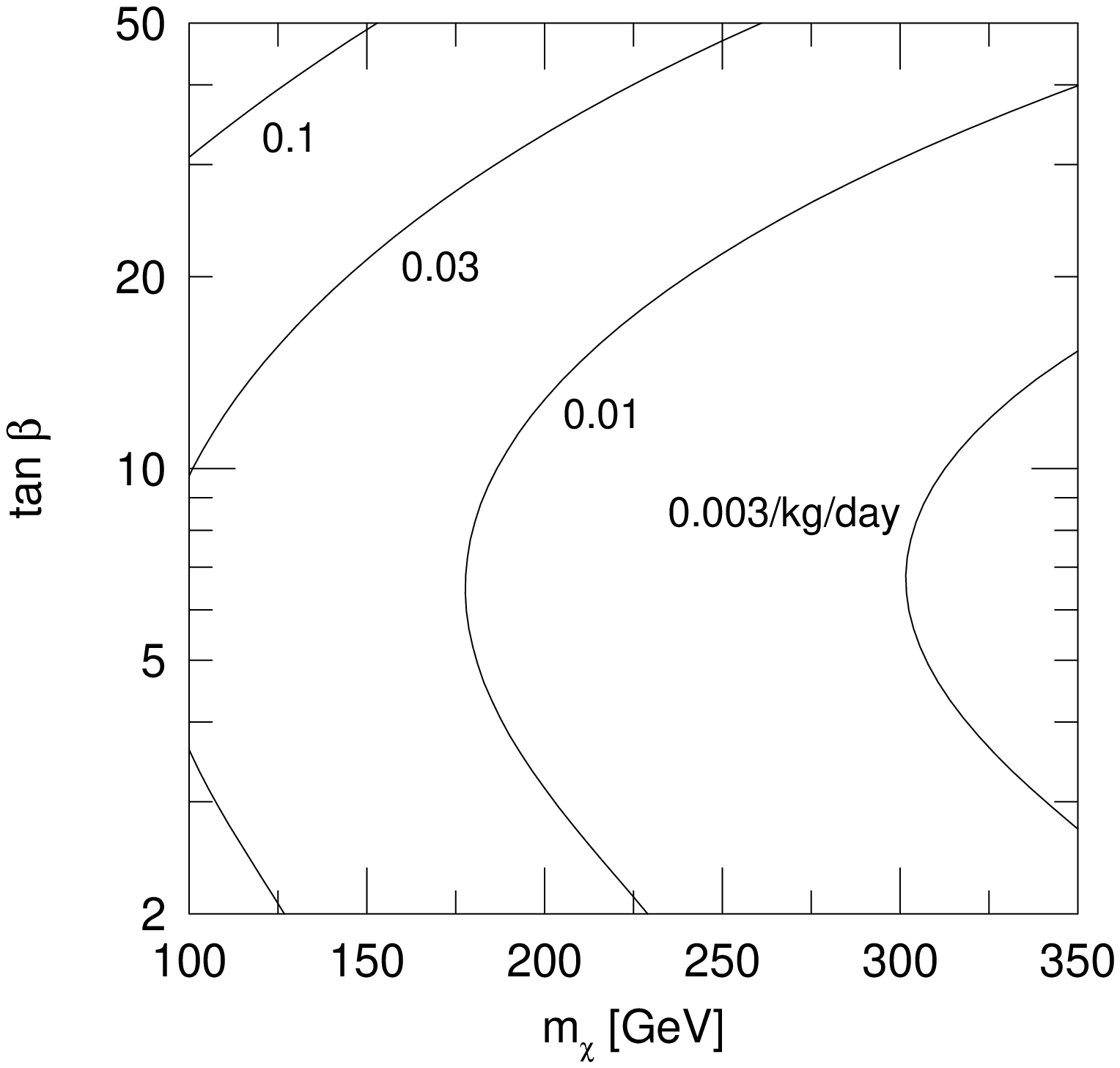}}
 \caption{Same as Fig.~\protect\ref{fig:R2}, except $\mu =3m_{\rm
G2}$.}
 \label{fig:R3}
 \end{figure}

In Figs.~\ref{fig:R2} and \ref{fig:R3}, we show the expected detection
rate of the Wino CDM in a $^{76}$Ge detector for $\mu =2m_{\rm G2}$
and $\mu= 3m_{\rm G2}$, respectively.  In our calculation, we used the
Higgs masses $m_h={\rm 100~GeV}$ and $m_H={\rm 300~GeV}$.  The other
parameters are taken to be $E_{\rm thr}={\rm 2~keV}$,
$\bar{v}_\chi={\rm 320~km/sec}$, and $\rho_\chi^{\rm (halo)}={\rm
0.3~GeV/cm^3}$.

The behavior of Figs.~\ref{fig:R2} and \ref{fig:R3} can be understood
as follows.  In the small $\tan\beta$ region, the scattering is
dominated by the light Higgs exchange diagram.  In this case, the
detection rate $R$ decreases as $\tan\beta$ increase since
$y_{h\chi\chi}$ is more suppressed for large $\tan\beta$ (if
$\mu/m_{\rm G2}>0$).  On the other hand, for the large $\tan\beta$
region, heavy Higgs exchange is the dominant contribution.  Since the
$HNN$ coupling $y_{HNN}$ is proportional to $\tan\beta$, $R$ is more
enhanced for larger $\tan\beta$.  Furthermore, $R$ decreases as
$m_\chi\simeq |m_{\rm G2}|$ or $|\mu|$ increases, since the
Wino-Higgsino mixing is more suppressed in this limit.

We can see that the detection rate of order 0.1 $-$ 0.01~event/kg/day
is possible in a Ge detector, which is within the reach of the
on-going CDM searches~\cite{PRep267-195}.  The detection rate is
considerably larger than the conventional Bino CDM case.  This is
because the couplings $y_{h\chi\chi}$ and $y_{H\chi\chi}$ are
(approximately) proportional to $g_2^2$ for the Wino LSP, instead of
$g_1^2$ for the Bino LSP.  Indeed, in the minimal supergravity model
with the Bino LSP, the detection rate is typically $O(10^{-3}~{\rm
event/kg/day})$ or less~\cite{PRD48-3483}, which is an order of
magnitude smaller than what is detectable.\footnote
 {In gravity-mediated SUSY breaking with the GUT relation among the
gaugino mass parameters, a large Higgsino component in the LSP and/or
a non-universal boundary condition for the scalar masses would be
necessary for a detectable SUSY dark matter, unless $\tan\beta$ is
large. It is very unlikely that conventional SUSY dark matter will be
detected. }
 Therefore, the Wino CDM has more chance to be detected in a Ge
detector. Of course, the detection rate given in Eq.~(\ref{Rate}) is
sensitive to the model parameters.  For example, the discovery of the
signal will be difficult if $|\mu| \gg |m_{\rm G2}|$.  Furthermore, in
the large $\tan\beta$ region, $R\propto m_H^{-4}$ and the detection
rate decreases as $m_H$ increases.  However, too large $\mu$ and too
heavy Higgs are not preferred from the naturalness point of view.
Therefore, in a significant fraction of the parameter space of the
AMSB scenario, the Wino CDM should be detectable.

For $\mu/m_{\rm G2}<0$, $y_{h\chi\chi}$ is suppressed because of the
cancellation, as shown in Eq.~(\ref{y_h1xx}).  Therefore, in this
case, the detection of the Wino CDM is difficult for small
$\tan\beta$.  On the other hand, for large $\tan\beta$, the detection
rate is less sensitive to the relative sign between $\mu$ and $m_{\rm
G2}$, and detection might be possible.

There is an alternative means to search for dark matter, which is to
look for the energetic neutrinos produced by the annihilation of the
Wino LSP's captured in the center of the sun and/or earth~\cite{ID}.
Such a high energy neutrino can be converted to a high energy muon
which can be observed as an upward going muon in \v{C}erenkov
detectors.  Using the conversion factors given in
Ref.~\cite{PRL74-5174}, we find that an LSP with $R\sim 0.1-0.01~{\rm
event/kg/day}$ has the event rate for indirect detection in the
\v{C}erenkov detectors of $\Gamma_{\rm ID}\sim 10-1~{\rm
event/m^2/yr}$. This is less sensitive than the direct detection in a
Ge detector~\cite{PRL74-5174}.

It might be that the most promising method for searching for dark
matter is to look for anti-matter, either
anti-protons~\cite{PRL53-624} or
positrons~\cite{PRD42-1001,PRD43-1774}, produced by the pair
annihilation of the LSP in our galaxy.  The pair annihilation rate
$\langle v_{\rm rel}\sigma\rangle$ for the Wino LSP is given in
Eq.~(\ref{vs_WW}), and numerically is given by $3.8\times
10^{-24}~{\rm cm^3/sec}$ (for $m_\chi ={\rm 100~GeV}$) $-$ $0.9\times
10^{-24}~{\rm cm^3/sec}$ (for $m_\chi ={\rm 300~GeV}$).  Unlike
standard SUSY dark matter, this rate is very large, and we expect a
high flux of anti-particles.  Furthermore, there are several on-going
projects for measuring the anti-matter flux in the cosmic ray.  In
particular, a very accurate measurement is expected by the AMS
experiment~\cite{ams}, which is a search for anti-matters with ``Alpha
Magnetic Spectrometer'' on the space shuttle and on the international
space station.  Since the experiment is not affected by the
atmosphere, AMS will greatly improve the measurements of the
anti-matter fluxes in the cosmic ray.

A recent calculation of the anti-proton flux $\Phi_{\bar{p}}$ can be
found in Ref.~\cite{astro9902012} where the flux from an LSP which
dominantly annihilates into $W$-boson pair is presented (see Example
No.~4 in Ref.~\cite{astro9902012}).  Since $\Phi_{\bar{p}}$ from the
LSP annihilation is proportional to $\langle v_{\rm
rel}\sigma\rangle\times (\rho_\chi^{\rm (halo)}/m_\chi)^2$, we
estimate the anti-proton flux by rescaling the result given in
Ref.~\cite{astro9902012}.  Adopting the canonical astrophysical
parameters of Ref.~\cite{astro9902012}, we find $\Phi_{\bar{p}}$ to be
$0.46~{\rm m^{-2}sec^{-1}sr^{-1}GeV^{-1}}$ (for $m_\chi ={\rm
100~GeV}$) $-$ $0.012~{\rm m^{-2}sec^{-1}sr^{-1}GeV^{-1}}$ (for
$m_\chi ={\rm 300~GeV}$).  Experimentally, the anti-proton flux is
already measured by the BESS experiment, and is given by
$(1.36^{+0.86}_{-0.61})\times 10^{-2}~{\rm
m^{-2}sec^{-1}sr^{-1}GeV^{-1}}$~\cite{PRL81-4052}.  As a result, a
naive comparison of our estimate with the measured flux would already
imply the constraint $m_\chi\gtrsim 250~{\rm GeV}$.  One should note,
however, that the flux is proportional to ${\rho^{\rm
(halo)}_\chi}^2$.  Furthermore, if we change the model of the halo, it
affects the propagation of the anti-proton and the flux can be reduced
significantly~\cite{astro9902012}.  Therefore, the theoretical result
is very sensitive to astrophysics parameters, and we do not draw any
strong conclusion here.  It is unlikely that the situation for
anti-protons will improve with the AMS experiment, since the
anti-proton signal falls with higher energy faster than the
background. However, the situation for positrons is very promising as
we now discuss.

 \begin{table}[t]
 \begin{center}
 \begin{tabular}{ccc}
 \hline\hline
 {$m_\chi$}
 & {$\Phi_{e^+}/(\Phi_{e^-}+\Phi_{e^+})_{\rm Wino}$}
 & {$\Phi_{e^+}/(\Phi_{e^-}+\Phi_{e^+})_{\rm BG}$} \\
 \hline
 {100~GeV} & {0.15}  & {0.032} \\
 {300~GeV} & {0.049} & {0.026} \\
 \hline\hline
  \end{tabular}
 \caption{The positron fraction $\Phi_{e^+}/(\Phi_{e^-}+\Phi_{e^+})$
in the cosmic ray at the peak of the spectrum.  The second column is
the result with the Wino CDM.  The third column is the background
estimated from the conventional sources.}
 \label{table:e+}
 \end{center}
 \end{table}

In Ref.~\cite{astro9902012}, the positron flux, $\Phi_{e^+}$, is also
presented.  Rescaling the given result, we found that $\Phi_{e^+}$
from the Wino-like CDM is comparable to the currently measured flux by
the HEAT experiment~\cite{astro9712324} even for $m_\chi\simeq {\rm
100~GeV}$.  Therefore, we believe our scenario is not seriously
constrained by the present data, but could give a very prominent
signal in the future.  One important point is that the positron
spectrum from the Wino CDM is peaked at high energy ($\sim
\frac{1}{2}m_\chi$)~\cite{PRD42-1001,PRD43-1774}, which can be a
distinctive signature of the Wino annihilation in the halo.  In order
to study the significance of this signal, we rescaled the results
given in Ref.~\cite{PRD43-1774} to obtain the positron fraction
$\Phi_{e^+}/(\Phi_{e^-}+\Phi_{e^+})$ at the peak.\footnote
 {In Ref.~\cite{PRD43-1774}, positron fraction is calculated for the
Higgsino-like CDM which also dominantly annihilates into $W$- and
$Z$-bosons.  We estimated the positron fraction for the Wino-like CDM
by rescaling with Eq.~(\ref{vs_WW}).  We used the halo mass density to
be $\rho_\chi^{\rm (halo)}={\rm 0.3~GeV/cm^3}$.}
 The results are given in Table~\ref{table:e+}.  As one can see, with
the Wino CDM, the positron fraction can be about 5 $-$ 2 times larger
than the background for $m_\chi={\rm 100-300~GeV}$.  A measurement of
the positron fraction has recently performed at lower energies by AMS
who will also do very precise measurement at higher energies in the
future.  Since the signal can be peaked at higher energy, this should
be an excellent way to search for Wino CDM.\footnote
 {In fact, Ref.~\cite{astro9902162} suggests a distortion in the
positron spectrum measured by the HEAT experiment~\cite{astro9712324}.
Positrons from the Wino CDM may be the source of this distortion.}
The signal is much stronger for Wino CDM than the more frequently
studied Bino dark matter, which would be undetectable. The distortion
suggested here should provide strong evidence for the Wino CDM
scenario.

\section{Conclusion} 
\label{sec:discussion}
\setcounter{equation}{0}

In this paper, we discussed the cosmological moduli problem in the
sequestered sector/AMSB scenario.  In this scheme, the gravitino mass
(corresponding to the moduli mass) is naturally 10 $-$ 100~TeV.  As a
result, cosmological moduli fields can decay before BBN starts.
Furthermore, since the LSP is likely to be the Wino-like neutralino,
the production of the LSP through moduli decay can be suppressed and
the Universe should not be overclosed, contrary to the case of the
Bino-like LSP.  Moreover, the mass density of the Wino-like LSP can be
naturally close to the critical density of the Universe, and hence the
Wino is an interesting candidate for CDM.  Furthermore, we have seen
that if the halo density is dominated by the Wino CDM, the detection
rate of the Wino CDM in Ge detector can be as large as $0.1 -
0.01$~event/kg/day, which is within the reach of the CDM search
experiment, and furthermore, the positron signal in AMS should be
measurable.

\section*{Acknowledgements}

We would like to thank Ann Nelson, Paul Schechter, Yael Shadmi, and
Frank Wilczek for useful discussions.  This work was supported in part
by the National Science Foundation under contract NSF-PHY-9513835, and
in part by the Department of Energy under contracts DE-FG02-90ER40542,
DE-FG-02-91ER40671, and cooperative agreement DF-FC02-94ER40818.  The
work of TM was also supported by the Marvin L. Goldberger Membership.

\appendix

\section{Moduli Field Couplings}
\label{app:property}
\setcounter{equation}{0}

In this Appendix, we discuss the mass and couplings of the moduli
fields.

We assume that a modulus field $\phi$ acquires its mass from SUSY
breaking effects.  Therefore, its mass is expected to be of the order
of the gravitino mass $m_{3/2}\sim {\rm
10-100~TeV}$~\cite{PLB318-447}.  Some non-perturbative effects may be
able to give much larger masses to the moduli fields.  Such a heavy
modulus field is cosmologically safe since its lifetime can be much
shorter than 1~sec.

It is also important to understand the relevant operators which
contribute to the decay of $\phi$.  By taking the minimal SUSY
standard model (MSSM) as the low energy effective theory, the
following operators can exist.

The modulus field can decay into a gauge field (and gaugino) through
the operators:
 \begin{eqnarray}
  {\cal L}_{\rm G} = \int d^2 \theta
  \frac{\lambda_{\rm G}}{M_*}
  \phi W^\alpha W_\alpha + \mbox{h.c.},
 \label{L_G}
 \end{eqnarray}
 where $\lambda_{\rm G}$ is a constant to parameterize the strength of
this interaction. When the gaugino mass $m_{\tilde{G}}$ is much
smaller than $m_\phi$, the branching ratio for the decay into the
gaugino pair receives chirality suppression~\cite{PRL75-398}. As a
result, the (partial) decay width and $\bar{N}_{\rm LSP}$ is given by
 \begin{eqnarray}
  \Gamma_{\rm G} &=& \frac{N_{\rm f}\lambda_{\rm G}^2}{8\pi}
  \frac{m_\phi^3}{M_*^2},
 \\
  \bar{N}_{\rm LSP} &\sim& O(m_{\tilde{G}}^2 / m_\phi^2),
 \end{eqnarray}
 where $N_{\rm f}$ is the number of the possible final states. (For
example, $N_{\rm f}=N^2-1$ for an SU($N$) gauge group.) One should
note that this operator also contributes to the gaugino mass if the
$\phi$ field participates in SUSY breaking.  So either $\lambda_G$ is
suppressed or $F_\phi$ is in order to maintain the anomaly-mediated
predictions.

In the MSSM, the following operator is also allowed:
 \begin{eqnarray}
  {\cal L}_{\rm H} = \int d^4 \theta
   \frac{\lambda_{\rm H}}{M_*} \phi H_1^* H_2^* + \mbox{h.c.},
 \label{L_H}
 \end{eqnarray}
 where $H_1$ and $H_2$ are the down-type and up-type Higgses,
respectively.  With this operator, the modulus field can decay into a
Higgs boson pair, and the decay width for this process is
 \begin{eqnarray}
  \Gamma_{\rm H} = \frac{\lambda_{\rm H}^2}{8\pi}
  \frac{m_\phi^3}{M_*^2}.
 \end{eqnarray}
 Notice that the Higgsino cannot be produced from this operator, and
hence $\bar{N}_{\rm LSP}=0$ for this process. This operator is also
dangerous if $F_\phi$ is maximal since it generates too large
$\mu$-parameter.

One can also write down the following operator:
 \begin{eqnarray}
  {\cal L}_{\rm Q^*Q} = \int d^4 \theta
  \frac{\lambda_{\rm Q^*Q}}{M_*} \phi Q^* Q  + \mbox{h.c.}
  = \frac{\lambda_{\rm Q^*Q}}{M_*}
  \phi (\partial^2 \tilde{Q}^*) \tilde{Q} + \cdots,
 \end{eqnarray}
 where $Q$ is general chiral superfields in the MSSM and $\tilde{Q}$
is its scalar component.  As suggested from the structure of the
operator, the decay rate is given by
 \begin{eqnarray}
  \Gamma_{Q^*Q} \sim O \left(
  \frac{\lambda_{\rm Q^*Q}^2}{4\pi}
  \frac{m_{\tilde{Q}}^4}{m_\phi^4}
  \frac{m_\phi^3}{M_*^2}
  \right),
 \end{eqnarray}
 where $m_{\tilde{Q}}$ is the soft SUSY breaking mass for $\tilde{Q}$.
Therefore, the decay through this operator is suppressed by a factor
of $O(m_{\tilde{Q}}^4/m_\phi^4)$.

In supergravity, we also expect the following interaction:
 \begin{eqnarray}
  {\cal L}_{\psi} = - \frac{1}{M_*^2}
  e^{K/2} W \psi_\mu \sigma^{\mu\nu} \psi_\nu + \mbox{h.c.},
 \label{L_psi}
 \end{eqnarray}
 where $K$ and $W$ are the K\"ahler potential and superpotential, and
$\psi_\mu$ is the gravitino field, respectively.  If the decay
$\phi\rightarrow\psi_\mu\psi_\mu$ is kinematically allowed, the effect
of this operator can be important.  In this case, $\phi$ decays into a
gravitino pair; then the produced gravitino decays into the standard
model particle and its superpartner through the supercurrent
interaction. For this process, the effective decay width (i.e., the
inverse of the time scale of this process) and $\bar{N}_{\rm LSP}$ are
estimated as
 \begin{eqnarray}
  \Gamma_{\psi} &\sim& O
  \left( \frac{1}{4\pi} \frac{m_{3/2}^3}{M_*^2} \right),
 \\
  \bar{N}_{\rm LSP} &\sim& O(1).
 \end{eqnarray}

There are also several operators that permit three-body decay. For
example, a modulus field can couple to the Yukawa-type operator in the
$F$-term like\footnote
 {We assume the coupling of the modulus field to the light quarks in
$F$-term is more suppressed. Otherwise, large vacuum expectation value
of $\langle\phi\rangle\sim M_*$ induces too large Yukawa coupling
constants for light quarks.}
 \begin{eqnarray}
  {\cal L}_{\rm Y} = \int d^2 \theta \frac{\lambda_{\rm Y}}{M_*}
  \phi H_2 t_{\rm R}^{\rm c} q_{\rm L}
   + \mbox{h.c.},
 \label{L_3body}
 \end{eqnarray}
 where $H_2$, $t_{\rm R}^{\rm c}$, and $q_{\rm L}$ are chiral
superfields for up-type Higgs, right-handed top quark, and left-handed
third generation quark doublet, respectively.  With this operator, the
decay rate is calculated as
 \begin{eqnarray}
  \Gamma_{\rm Y} = \frac{3\lambda_{\rm Y}^2}{128\pi^3}
  \frac{m_\phi^3}{M_*^2}.
 \end{eqnarray}
 This is a three-body process, and hence the decay rate is more
suppressed than those of the two-body processes.  For this process,
$\bar{N}_{\rm LSP}\sim O(1)$. One may also write down the scalar
interaction of the form $\frac{m_{3/2}}{M_*}\phi H_2\tilde{t}_{\rm
R}^{\rm c}\tilde{q}_{\rm L}$ arising from the supergravity effect. If
the coefficient of this operator is $O(1)$, however, too large an
$A$-parameter is generated with $\langle\phi\rangle\sim M_*$.
Therefore, we assume this operator is somehow suppressed if
$\langle\phi\rangle\sim M_*$.

So, to summarize, the moduli fields are expected to have a decay rate
as large as that estimated on dimensional grounds. The operators that
could lead to this decay rate are the operator given in
Eq.~(\ref{L_psi}) which leads to the decay to gravitinos (if
kinematically permitted), the coupling to gauge pairs (\ref{L_G}), and
the coupling to Higgses (\ref{L_H}).  However, a large value for the
last two couplings require a small value for $F_\phi$.  The decay rate
with no suppression factors is consistent with the requirements for
avoiding the cosmological moduli problem.

The density of Wino dark matter depends on $\bar{N}_{\rm LSP}$.  In
general, $\bar{N}_{\rm LSP}$ is expected to be small. The precise
value depends on the magnitude of the coupling of the associated
operators, but values of order $10^{-4}-10^{-2}$ are expected. The
exception to this small value is the operator that permits decay to
gravitinos. If kinematically possible, this would permit large
$\bar{N}_{\rm LSP}\sim 1$.


\begin{thebibliography}{99}

\bibitem{moduli_prob}
  G.D.~Coughlan, W.~Fischler, E.W.~Kolb, S.~Raby and G.G.~Ross,
    {\sl Phys.~Lett.} {\bf B131} (1983) 59;
  T.~Banks, D.B.~Kaplan and A.E.~Nelson,
    {\sl Phys.~Rev.} {\bf D49} (1994) 779;
  T.~Banks, M.~Berkooz and P.J.~Steinhardt,
    {\sl Phys.~Rev.} {\bf D52} (1995) 705.

\bibitem{PLB318-447}
  B.~de~Carlos, J.A.~Casas, F.~Quevedo and E.~Roulet,
    {\sl Phys.~Lett.} {\bf B318} (1993) 447.

\bibitem{PLB174-176}
  J.~Ellis, D.V.~Nanopoulos and M.~Quiros,
    {\sl Phys.~Lett.} {\bf B174} (1986) 176.

\bibitem{PLB342-105}
  T.~Moroi, M.~Yamaguchi and T.~Yanagida,
    {\sl Phys.~Lett.} {\bf B342} (1995) 105.

\bibitem{NPB449-229}
  L.~Randall and S.~Thomas,
    {\sl Nucl.~Phys.} {\bf B449} (1995) 229.

\bibitem{PLB370-52}
  M.~Kawasaki, T.~Moroi and  T.~Yanagida,
    {\sl Phys.~Lett.} {\bf B370} (1996) 52.

\bibitem{PLB438-267}
  T.~Nagano and M.~Yamaguchi,
    {\sl Phys.~Lett.} {\bf B438} (1998) 267.

\bibitem{hth9810155}
  L.~Randall and R.~Sundrum,
    hep-th/9810155.

\bibitem{JHEP9812-027}
  G.~Giudice, M.~Luty, H.~Murayama and R.~Rattazzi,
    {\sl JHEP} {\bf 9812} (1998) 027.

\bibitem{PRL75-398}
  M.~Dine, L.~Randall and S.~Thomas,
    {\sl Phys.~Rev.~Lett.} {\bf 75} (1995) 398.

\bibitem{lateinf}
  D.H. Lyth and E.D. Stewart,
    {\sl Phys.~Rev.~Lett.} {\bf 75} (1995) 201;
    {\sl Phys.~Rev.} {\bf D53} (1996) 1784.

\bibitem{hph9904250}
  J.L.~Feng, T.~Moroi, L.~Randall, M.~Strassler and S.~Su,
    hep-ph/9904250.

\bibitem{hph9904378}
  T.~Gherghetta, G.F.~Giudice and J.D.~Wells,
    hep-ph/9904378.

\bibitem{Kolb-Turner}
  See, for example, E.W. Kolb and M.S. Turner,
    {\sl The Early Universe} (Addison-Wesley, 1990).

\bibitem{PRL82-4168}
  M.~Kawasaki, K.~Kohri and N.~Sugiyama,
    {\sl Phys.~Rev.~Lett.} {\bf 82} (1999) 4168.

\bibitem{PRL48-1303}
  S.~Weinberg,
    {\sl Phys.~Rev.~Lett.} {\bf 48} (1982) 1303.

\bibitem{grav_bbn}
  M.Y.~Khlopov and A.D.~Linde,
    {\sl Phys.~Lett.} {\bf B138} (1984) 265;
  J.~Ellis, E.~Kim and D.V.~Nanopoulos,
    {\sl Phys.~Lett.} {\bf B145} (1984) 181.

\bibitem{PTP93-879}
  M.~Kawasaki and T.~Moroi,
    {\sl Prog.~Theor.~Phys.} {\bf 93} (1995) 879.

\bibitem{NPB249-361}
  I.~Affleck and M.~Dine,
    {\sl Nucl.~Phys.} {\bf B249} (1985) 361.

\bibitem{PRep267-195}
  See, for example, G.~Jungman, M.~Kamionkowski and K.~Griest,
    {\sl Phys. Rept.} {\bf 267} (1996) 195.

\bibitem{PLB78-443}
  M.A.~Shifman, A.I.~Vainstein and V.I.~Zakharov,
    {\sl Phys.~Lett.} {\bf B78} (1978) 443.

\bibitem{PRD48-3483}
  M.~Drees and M.M.~Nojiri,
    {\sl Phys.~Rev.} {\bf D48} (1993) 4483.

\bibitem{ID}
  J.~Silk, K.~Olive and M.~Srednicki,
    {\sl Phys.~Rev.~Lett.} {\bf 55} (1985) 257;
  L.M.~Krauss, M.~Srednicki and F.~Wilczek,
    {\sl Phys.~Rev.} {\bf D33} (1986) 2079;
  K.~Freese,
    {\sl Phys.~Lett.} {\bf B167} (1986) 295.

\bibitem{PRL74-5174}
  M.~Kamionkowski, K.~Griest, G.~Jungman and B.~Sadoulet,
    {\sl Phys.~Rev.~Lett.} {\bf 74} (1995) 5174.

\bibitem{PRL53-624}
  J.~Silk and M.~Srednicki,
    {\sl Phys.~Rev.~Lett.} {\bf 53} (1984) 624.

\bibitem{PRD42-1001}
  M.S.~Turner and F.~Wilczek,
    {\sl Phys.~Rev.} {\bf D42} (1990) 1001.

\bibitem{PRD43-1774}
  M.~Kamionkowski and M.S.~Turner,
    {\sl Phys.~Rev.} {\bf D43} (1991) 1774.

\bibitem{ams}
  The AMS collaboration,
    {\tt http://hpl3tri1.cern.ch/}.

\bibitem{astro9902012}
  L.~Bergstr\"{o}m, J.~Edsj\"{o} and P.~Ullio,
    astro-ph/9902012.

\bibitem{PRL81-4052}
  H.~Matsunaga {\sl et al.},
    {\sl Phys.~Rev.~Lett.} {\bf 81} (1998) 4052.

\bibitem{astro9712324}
  S.W.~Barwick {\sl et al.},
    astro-ph/9712324.

\bibitem{astro9902162}
  S.~Coutu {\sl et al.},
    astro-ph/9902162.



\end{thebibliography}
\end{document}